\documentclass[prl,nofootinbib,twocolumn,superscriptaddress]{revtex4}
\pdfoutput=1

\usepackage{graphicx,rotating}
\usepackage{hyperref}
\usepackage{ifpdf}
\usepackage[english]{babel}
\usepackage{amsmath}
\usepackage{amssymb}

\newcommand{\mrm}[1]{\mbox{\rm #1}}
\newcommand{\be}{\begin{equation}}
\newcommand{\ee}{\end{equation}}
\newcommand{\br}{\begin{eqnarray}}
\newcommand{\bea}{\begin{eqnarray}}
\newcommand{\eea}{\end{eqnarray}}
\newcommand{\er}{\end{eqnarray}}
\newcommand{\ba}{\begin{array}}
\newcommand{\ea}{\end{array}}
\newcommand{\bi}{\begin{itemize}}
\newcommand{\ei}{\end{itemize}}
\newcommand{\bn}{\begin{enumerate}}
\newcommand{\en}{\end{enumerate}}
\newcommand{\bc}{\begin{center}}
\newcommand{\ec}{\end{center}}

\newcommand{\Eq}[1]{Eq.~(\ref{#1})}
\newcommand{\rfn}[1]{(\ref{#1})}

\newcommand{\gsim}{\lower.7ex\hbox{$\;\stackrel{\textstyle>}{\sim}\;$}}
\newcommand{\lsim}{\lower.7ex\hbox{$\;\stackrel{\textstyle<}{\sim}\;$}}

\newcommand{\hc}[1]{#1^{\dagger}} % Hermitian conjugate

\newcommand{\Veff}{V_{\text{eff}}}

\def\mysection#1{{\bf #1.} }

\begin{document}
%\draft

%\preprint

\title{\bf 
EWSB from the soft portal into Dark Matter and prediction for direct detection
%EWSB from the soft portal into Dark Matter and the possibility for direct detection at CDMS II
}

\author{ Mario Kadastik}
\author{ Kristjan Kannike}
\author{ Antonio Racioppi}
\affiliation{National Institute of Chemical Physics and Biophysics, 
Ravala 10, Tallinn 10143, Estonia
} 

\author{ Martti Raidal}

\affiliation{National Institute of Chemical Physics and Biophysics, 
Ravala 10, Tallinn 10143, Estonia
} 
\affiliation{Department of Physics, P.O.Box 64, FIN-00014 University of Helsinki, Finland
}

%\date{\today}
%\pacs{}

\vspace*{-2.0 in}

\begin{abstract}
Scalar Dark Matter (DM) can have dimensionful coupling to the Higgs boson -- the ``soft" portal into DM -- 
which is predicted to be unsuppressed  by underlying $SO(10)$ GUT. The dimensionful coupling  can be large,
$\mu/v \gg 1,$ without spoiling perturbativity of  low energy theory up to the GUT scale. We show that the soft portal
into DM  naturally triggers radiative EWSB via large 1-loop DM corrections to the effective potential.
 In this scenario EWSB, DM thermal freeze-out cross section and DM scattering on nuclei are all dominated by the same coupling,
predicting DM mass range to be $700~\mrm{GeV}< M_{\rm{DM}} < 2~\mrm{TeV}.$ 
 The spin-independent direct detection cross section is predicted to be
just at the present experimental sensitivity and can explain the observed CDMS II recoil events.

\end{abstract}

%\vskip -3.5cm

\maketitle

\mysection{Introduction}
The aim of Tevatron and LHC experiments is to reveal the origin of electroweak symmetry breaking (EWSB).
In the standard model (SM) EWSB occurs spontaneously due to the negative Higgs boson mass parameter
$\mu^2_1$ in the scalar potential. Although negative $\mu^2_1$ is theoretically consistent in the SM, 
it prevents embedding  the SM into  grand unified theories~\cite{Georgi:1974sy,Fritzsch:1974nn}  (GUTs) 
in such a way  that unbroken SM gauge symmetry $SU(2)_L\times U(1)_Y$ emerges from GUT symmetry breaking.
Therefore the GUT paradigm invites for {\it dynamical} mechanisms for EWSB due to particle interactions. 
The most well known attempt to this direction is EWSB due to radiative corrections to the effective potential~\cite{Coleman:1973jx}.

The possibility that  cold Dark Matter (DM) of the Universe~\cite{Komatsu:2008hk} consists of
$Z_2$-odd SM singlet~\cite{rs} or doublet~\cite{id}  scalars is well motivated. 
Ultraviolet completion of the low scale scalar DM models is non-SUSY $SO(10)$ GUT which 
{\it predicts} scalar singlet/doublet DM~\cite{Kadastik:2009cu} due to the GUT-generated matter parity~\cite{Kadastik:2009dj}.
The scalar models  allow  to relate thermal  DM  properties to EWSB via scalar potential parameters.
Unfortunately, to achieve  radiative EWSB  due to 
the DM scalar doublet~\cite{Hambye:2007vf} or  
singlet~\cite{Gildener:1976ih,Hempfling:1996ht,Meissner:2006zh,Espinosa:2007qk,Chang:2007ki,Foot:2007as,Iso:2009ss}, 
some 
scalar quartic couplings must be very large, $\lambda_i \sim {\cal O}(4\pi),$ to overcome large opposite contribution
of 12 top quark {\it d.o.f.}
Such scenarios can be regarded as low energy models of EWSB but they cannot originate directly from GUTs
because the existence of  ${\cal O}(1)$~TeV scale Landau poles for $\lambda_i .$

Unlike the SM gauge and fermion sectors,  properties of the fundamental scalar sector are not known.
This offers unique possibilities for scenarios of new physics.
It was re-emphasized recently~\cite{Patt:2006fw}  that the SM Higgs boson mass term $\mu_1^2 H_1^\dagger H_1,$ 
being superrenormalizable, opens a portal into hidden sector via $d=4$ strictly renormalizable scalar operators of a
 form $\lambda_\Phi \Phi^\dagger \Phi  H_1^\dagger H_1,$ where $\Phi$ is a hidden sector scalar.
The scalar DM models~\cite{rs,id},  the radiative EWSB 
models~\cite{Gildener:1976ih,Hempfling:1996ht,Meissner:2006zh,Espinosa:2007qk,Chang:2007ki,Foot:2007as,Hambye:2007vf,Iso:2009ss}, the models of modified Higgs boson phenomenology~\cite{Schabinger:2005ei}, like
the ``hidden valley'' models~\cite{Strassler:2006im}, all belong to this class of {\it ``hard" portal}  because the
interaction  $\lambda_\Phi$ is dimensionless.

The scalar sector is unique among all other fundamental interactions that it also allows for superrenormalizable  {\it dimensionful interactions.}
The form of {\it ``soft" portal} discussed in~\cite{vanderBij:2006ne,Patt:2006fw}, $\mu S H_1^\dagger H_1,$ where $S$ is a SM 
singlet, cannot occur in case of scalar DM because it explicitly breaks $S\to -S.$ 
Model independently, the minimal Higgs boson {\it soft portal  into  DM}  is uniquely 
determined by the SM gauge symmetry  and  $H_{1} \to H_{1}$, $S \to -S,$ $ H_{2} \to -H_{2},$ 
\bea
V= \frac{\mu_{S H}}{2} \left[\hc{S} \hc{H_{1}} H_{2} + \mathrm{h.c.} \right]
+ \frac{\mu'_{S H}}{2} \left[S \hc{H_{1}} H_{2} + \mathrm{h.c.} \right], 
\label{si}
\eea
where $H_2$ is the inert doublet \cite{id} and $S$ is the complex singlet \cite{rs}. 
In the $SO(10)$ GUT  DM scenario~\cite{Kadastik:2009cu}, the interaction \rfn{si} is unsuppressed at low energies since the $\mu'_{S H}$ term is allowed by the $SO(10)$ gauge
symmetry. Because the interaction \rfn{si} is soft, it does not generate quadratic divergences to scalar masses
improving the hierarchy problem  arising from scalar self-interactions. 
Most importantly, the Landau poles do not appear in soft interactions and the model
remains perturbative all the way from $M_Z$ to the GUT scale $M_\text{G}$ even if $\mu'_{SH}/v \gg 1.$

In this work we study the minimal scalar DM model possessing the soft interaction portal \rfn{si} and require that 
the model remains perturbative up to the GUT scale. This implies that all $\mu^2_i(M_\text{DM})>0,$ 
all $\lambda_i(M_\text{DM}) \lsim {\cal O}(0.1),$   and
the only possible source of large scalar couplings is \Eq{si}. We compute renormalization group (RG) 
improved one-loop effective potential $V_{\text{eff}}$ of the multi-scale model characterized by 
$M_\text{G} \ggg M_{\text{DM}}\sim  \mu'_{SH}\gg M_h.$ 
Because the interaction \rfn{si} is dimensionful,  it induces large negative contribution only to the 
SM Higgs mass parameter $\mu^2_1.$
 Below the DM decoupling scale the {\it improved} SM Higgs boson scalar potential 
is of the minimal form, $V={\bar\mu}_{1}^{2} h^2 +\lambda_1 h^4,$ with radiatively induced negative ${\bar\mu}_{1}^{2}<0,$ 
possessing the minimum  for small field values $h=v\ll M_{\text{DM}}.$

To our best knowledge the proposed scenario is the first one which successfully applies the idea of radiative 
EWSB from effective potential using dimensionful interactions. 
Effective potentials with several explicit mass scales \cite{bando,Casas:1998cf}
are very different from the Coleman-Weinberg original proposal~\cite{Coleman:1973jx}
in which EWSB is induced without any explicit scale.
We propose that the EW scale is obtained from the DM scale by loop suppression,
opening a completely new way of understanding the EWSB.

 Phenomenology of the proposed scenario of DM and EWSB is very predictive.
 Because $\mu'_{SH}/v \gg 1,$ the radiative EWSB, the thermal freeze-out cross section of the 
lightest DM particle and the spin independent DM direct detection cross section with nuclei are {\it all}
dominated by the same soft interaction term $\mu'_{SH}.$ This implies that  we get 
{\it predictions} for the DM mass and the DM direct detection cross section from the measurements of the 
DM abundance of the Universe and the SM Higgs boson mass.
We show that the DM direct detection cross section is predicted to be just at the present experimental sensitivity
and can explain the observed DM recoil events at CDMS II \cite{CDMS}.
In this scenario the Higgs boson mass measurement at LHC essentially fixes the properties of DM.
 
We assume that the hierarchy problem is solved by (yet unknown) physics above the GUT scale, and below that
quantum field theory with $\overline{MS}$ renormalization scheme is an appropriate tool for computations.

\mysection{The origin of soft portal into DM}
Although our main result of radiative EWSB, based on \Eq{si}, is model independent, 
it is illustrative to work within the concrete model predicting \Eq{si} -- the non-SUSY SO(10) GUT~\cite{Kadastik:2009cu}.
The minimal model has one scalar ${\bf 10}$ representation containing the SM Higgs doublet $H_1$
and one scalar ${\bf 16}$ containing the  DM doublet $H_2$ and singlet $S.$
Below $M_\text{G}$ and above the EWSB scale
the model is described by the $H_{1} \to H_{1}$, $S \to -S,$ $ H_{2} \to -H_{2}$ 
invariant scalar tree-level potential
\begin{equation}
\begin{split}
V &= \mu_{1}^{2} \hc{H_{1}} H_{1} + \lambda_{1} (\hc{H_{1}} H_{1})^{2} 
+ \mu_{2}^{2} \hc{H_{2}} H_{2} + \lambda_{2} (\hc{H_{2}} H_{2})^{2} \\
&+ \mu_{S}^{2} \hc{S} S + \frac{\mu_{S}^{\prime 2}}{2} \left[ S^{2} + (\hc{S})^{2} \right] + \lambda_{S} (\hc{S} S)^{2} 
 \\
& + \frac{ \lambda'_{S} }{2} \left[ S^{4} + (\hc{S})^{4} \right] 
 + \frac{ \lambda''_{S} }{2} (\hc{S} S) \left[ S^{2} + (\hc{S})^{2} \right] \\
&+ \lambda_{S1}( \hc{S} S) (\hc{H_{1}} H_{1}) + \lambda_{S2} (\hc{S} S) (\hc{H_{2}} H_{2}) \\
&+ \frac{ \lambda'_{S1} }{2} (\hc{H_{1}} H_{1}) \left[ S^{2} + (\hc{S})^{2} \right]
+ \frac{ \lambda'_{S2} }{2} (\hc{H_{2}} H_{2}) \left[ S^{2} + (\hc{S})^{2} \right]
 \\
&+ \lambda_{3} (\hc{H_{1}} H_{1}) (\hc{H_{2}} H_{2}) + \lambda_{4} (\hc{H_{1}} H_{2}) (\hc{H_{2}} H_{1}) \\
&+ \frac{\lambda_{5}}{2} \left[(\hc{H_{1}} H_{2})^{2} + (\hc{H_{2}} H_{1})^{2} \right] \\
&+ \frac{\mu_{S H}}{2} \left[\hc{S} \hc{H_{1}} H_{2} + \mathrm{h.c.} \right]
+ \frac{\mu'_{S H}}{2} \left[S \hc{H_{1}} H_{2} + \mathrm{h.c.} \right], \nonumber
\end{split}
\end{equation}
together with the GUT scale boundary conditions
\bea
&\mu_1^2(M_{\text{G}})>0,\; \mu'_{SH}(M_\text{G})\neq 0,\; \mu_2^2(M_{\text{G}})=\mu_S^2(M_{\text{G}}) , & 
\label{bc1}\\
&\lambda_2(M_{\text{G}})=\lambda_S(M_{\text{G}})=\lambda_{S2}(M_{\text{G}}),\; \lambda_3(M_{\text{G}})=\lambda_{S1}(M_{\text{G}}), & \nonumber
\eea
and
\bea
&{\mu}_S^{\prime 2}, \; {\mu}_{SH}^{ 2} \lsim {\cal O}\left( \frac{M_\text{G}}{M_{\text{P}}}\right)^n \mu^2_{1,2},& 
\label{bc2}\\\
& \lambda_{5} ,\; \lambda'_{S1} ,\; \lambda'_{S2} ,\; \lambda''_{S} \lsim {\cal O}\left( \frac{M_\text{G}}{M_{\text{P}}}\right)^n \lambda_{1,2,3,4}.&
\nonumber
\eea
While the parameters in \Eq{bc1} are allowed by
$SO(10),$ the ones in \Eq{bc2} can be generated only after $SO(10)$ breaking by  operators
suppressed by $n$ powers of the Planck scale $M_{\text{P}}.$
Notably, the soft interaction term $\mu'_{SH}$ is allowed by $SO(10)$ and should be of the same
order as the other unsuppressed DM mass parameters $\mu_2,$ $\mu_S.$

\mysection{Improved effective potential with explicit mass scales}
Working in the $\overline{MS}$ renormalization scheme, the one-loop effective potential in the direction of the SM Higgs 
boson $h$ is
\begin{equation}
\begin{split}
\Veff (h) & = 
\frac{\mu_{1}^2}{2} h^{2} + \frac{\lambda_{1}}{4} h^{4} + 
 \frac{1}{64 \pi^{2}} \left[ m_{G_{0}}^4 \left( \ln \frac{m_{G_{0}}^2}{\mu^2} -\frac{3}{2} \right) \right. \\
& + 2 m_{G^{+}}^4 \left( \ln \frac{m_{G^{+}}^2}{\mu^2} -\frac{3}{2} \right) +  m_{h}^{4} \left(\ln \frac{m_{h}^{2}}{\mu^2} -\frac{3}{2} \right)  \\
 & + \sum_i^4  m_{i}^{4} \left(\ln \frac{m_{i}^{2}}{\mu^2} -\frac{3}{2} \right) +  2 m_{H^{+}}^{4} \left( \ln \frac{m_{H^{+}}^{2}}{\mu^2} -\frac{3}{2} \right) 
  \\ 
   &  +
    2 m_{Z}^{4} \left( \ln \frac{m_{Z}^{2}}{\mu^2} -\frac{5}{6} \right) +
   4 m_{W}^{4} \left( \ln \frac{m_{W}^{2}}{\mu^2} -\frac{5}{6} \right) \\
   & \left.  - 12 m_{t}^{4} \left( \ln \frac{m_{t}^{2}}{\mu^{2}} - \frac{3}{2} \right) \right],
\end{split}
\label{Veff0}
\end{equation}
where the field-dependent masses are $m_{h}^{2} = \mu_{1}^{2} + 3 \lambda_{1} h^{2}$, $m_{G_{0}} = m_{G^{+}} = \mu_{1}^{2} + \lambda_{1} h^{2}$, $m_{t} = y_{t}^{2} h^{2}/2$, $m_{Z}^{2} = (g^{2} + g^{\prime 2}) h^{2}/4,$  $m_{W}^{2} = g^{2} h^{2}/4$ and $m_{H^{+}}^{2} = \mu_{2}^{2} + \lambda_{3} h^{2}/2.$
The  neutral $Z_2$-odd scalar masses $m_{1,2}^{2}$ are obtained by diagonalization of the mass matrix
\begin{equation}
\begin{pmatrix}
\mu_{2}^{2}+ (\lambda_{3} + \lambda_{4} + \lambda_{5}) h^{2}/2 
& (\mu_{SH} + \mu'_{SH}) h/(2 \sqrt{2}) \\
 (\mu_{SH} + \mu'_{SH}) h/(2 \sqrt{2}) 
 & \mu_{S}^{2} + \mu_{S}^{\prime 2} + (\lambda_{S1} + \lambda'_{S1} ) h^2/2
 \end{pmatrix}. \nonumber
 \label{eq:M:S:real:pot}
\end{equation}
The pseudo-scalar masses $m_{3,4}^{2}$ are obtained from $m_{1,2}^{2}$ 
by replacing $\lambda_5 \to -\lambda_5,$ 
$\lambda'_{S1} \to -\lambda'_{S1},$  $ \mu_{S}^{\prime 2} \to  -\mu_{S}^{\prime 2}.$ 
Here and in the following we denote tree-level masses with small $m^{2}_{i}$ and 
the physical one-loop masses with $M^{2}_{i}$. Notice that the stability of the 
potential \rfn{Veff0} requires that the quartic couplings $\lambda_i$ are non-vanishing. The positivity of dark sector masses requires $\mu_{SH}^{\prime 2} v^{2}/2 < \, [2 \mu_{2}^{2} + (\lambda_{3} + \lambda_{4}) v^{2}]\, [2 \mu_{S}^{2} + \lambda_{S1} v^{2}],$ where the parameters  \rfn{bc2} are neglected.

\Eq{Veff0} depends explicitly on the renormalization scale $\mu.$ This is unphysical
because by definition $d \Veff/d\mu=0.$ 
In order to compute physical effects,  \Eq{Veff0} must be RG improved.
It was shown in Ref.~\cite{Casas:1998cf} that the correct principle to RG improve the potential having 
several explicit mass scales is the decoupling theorem \cite{Appelquist:1974tg}. In effect, each $m_{i}^{2}$ term in $V_{\text{eff}}$ is multiplied by the step function $\theta(\mu^{2} - m_{i}^{2}(h))$.
We fix the boundary conditions \rfn{bc1}, \rfn{bc2} at $M_\text{G}$ and run all the parameters 
 to the largest mass scale in \rfn{Veff0} calculated for a chosen field value $h$ 
using 1-loop RGEs of the full model~\cite{Kadastik:2009cu}. 
We evaluate the effective potential \rfn{Veff0}
at that scale using the renormalized parameters and decouple the dark sector particle.
Below the decoupling scale new low energy model is obtained by matching the
model parameters above and below decoupling. Repeating this procedure until all heavy particles are 
decoupled one obtains the desired RG improved effective potential for the given $h$.
Repeating the procedure for every field value $h$ gives the full $\Veff.$

\begin{figure}[t]
\includegraphics[width=0.37\textwidth]{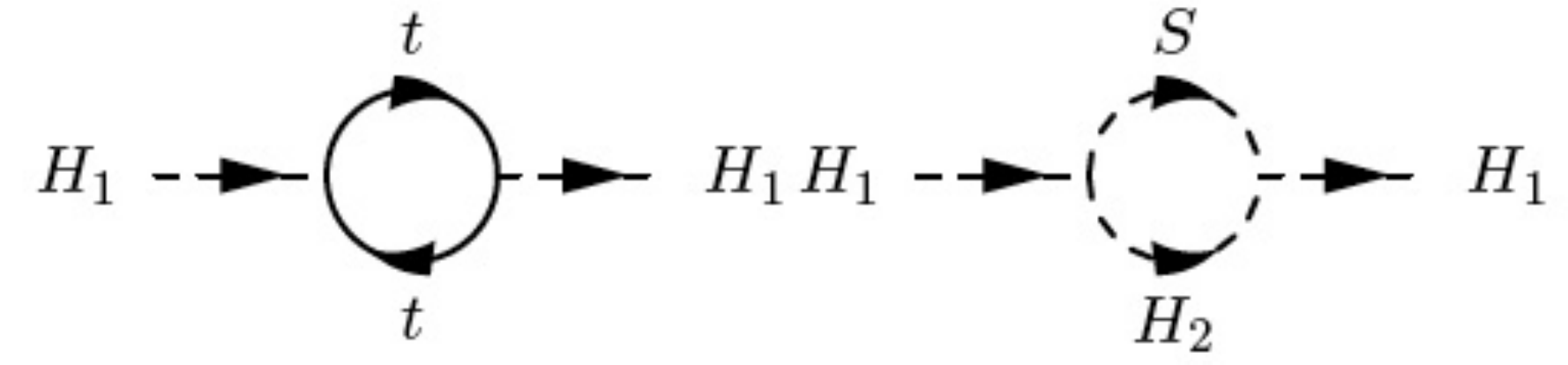}
\caption{Dominant 1-loop diagrams contributing to $\mu^2_1.$
}
\label{fig1}
\end{figure}

In our computation  we consider only the numerically dominant 1-loop corrections from top quark and the DM fields, 
depicted in Fig. \ref{fig1}. As explained above, the DM coupling in Fig. \ref{fig1} is dominated by the soft interaction $\mu'_{SH}.$ 
For simplicity, we decouple all the dark sector fields at the common scale $(m_{1} + m_{2})/2$. 
Below the DM decoupling scale $\Veff$ becomes
\begin{equation}
\Veff (h) = 
\frac{\bar{\mu}_{1}^2}{2} h^{2} + \frac{\bar{\lambda}_{1}}{4} h^{4} -
\frac{3}{16 \pi^{2}}  m_{t}^{4} \left( \ln \frac{m_{t}^{2}}{\mu^{2}} - \frac{3}{2} \right), 
\label{Veff1}
\end{equation}
where
\begin{equation}
  \bar{\mu}_{1}^{2} = \mu_{1}^{2} - \frac{3}{128 \pi^{2}}
  \left[\mu_{SH}^{\prime 2} + 4 (2 \lambda_{3} + \lambda_{4}) \mu_2^2 + 
    4 \lambda_{S1} \mu_S^2 \right],
\label{mucorr}
\end{equation}
and
\begin{equation}
  \bar{\lambda}_{1} = \lambda_{1} - \frac{3}{64 \pi^{2}}
  \left[(\lambda_{3} + \lambda_{4})^2 + \lambda_{3}^{2} + \lambda_{S1}^2 \right],
  \label{lamcorr}
\end{equation}
have been obtained by matching. All the parameters in \rfn{mucorr}, \rfn{lamcorr}
are the running quantities evaluated at the proper scale. Below the DM decoupling scale 
we use the SM RGEs to to run the parameters in \rfn{Veff1} to the top quark mass scale.
After decoupling the top quark  $\Veff$ becomes
\begin{equation}
\Veff (h) = 
\frac{\bar{\mu}_{1}^2}{2} h^{2} + \frac{\bar{\bar{\lambda}}_{1}}{4} h^{4},
\label{Veff2}
\end{equation}
where $\bar{\mu}_{1}^{2}$ is given by \eqref{mucorr} and $\bar{\bar{\lambda}}_{1} = \bar{\lambda}_{1} + 9 y_{t}^{4}/32 \pi^{2}.$

%Similar calculations of $\Veff$  in dark scalar directions  do not give qualitatively new  physics effects.

\mysection{EWSB} 
For large field values $h$ the scalar potential is clearly dominated by the quartic 
couplings. For small field values $h\lsim M_{\text{DM}}$
 the Higgs mass parameter $\mu_1^2,$ required to be {\bf positive} above $M_{\rm DM},$ 
 gets a large negative contribution \rfn{mucorr} from the 
DM loop  which is dominated by the $\mu^{\prime 2}_{SH}$ term.
The correction \rfn{lamcorr} is small because $\lambda_i\lsim 0.1.$  If the correction \rfn{mucorr} drives
${\bar\mu}^2_1$ negative, EWSB occurs radiatively due to low scale effects in the scalar potential.
The minimization condition $\text{d}\Veff/\text{d}h = 0$ implies $ M_{h}^{2} = -2 \bar{\mu}_{1}^2,$ as in the SM.
In this case all the parameters in \rfn{Veff2} are the low energy effective 1-loop RG improved parameters.
However, our result \Eq{mucorr} is general and, in the absence of a high energy theory, can be viewed as 
the low energy radiative EWSB mechanism.

\begin{figure}[t]
\includegraphics[width=0.37\textwidth]{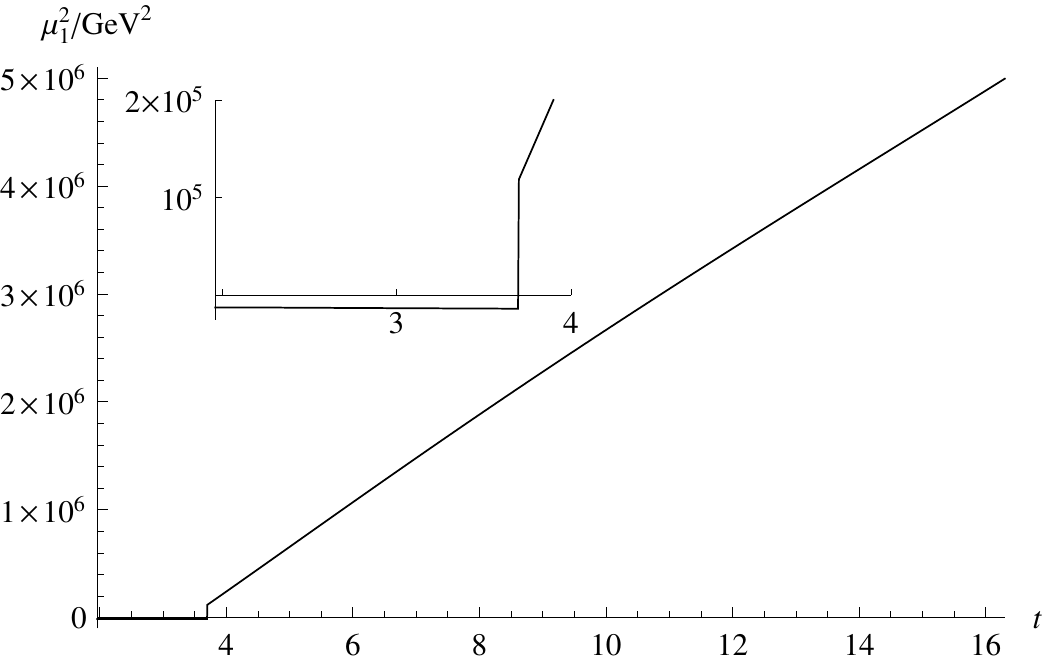}
\caption{An example of $\mu_1$ running from $M_\text{G}$ to $M_Z$ ($t = \log \mu/\text{GeV}$).  
The inset enlarges the low energy region, demonstrating EWSB due to 1-loop corrections from DM. 
}
\label{fig2}
\end{figure}

In Fig.~\ref{fig2} we plot an example of the evolution of $\mu^2_1$ from $M_\text{G}$ to $M_Z.$
Although $\mu^2_1$ decreases with energy due to the RGE effects, it remains positive at $M_{\text{DM}}.$
The DM 1-loop effect to the effective potential, enlarged in the inset of Fig.~\ref{fig2}, drives the mass parameter negative 
according to \Eq{mucorr} and triggers the EWSB. The origin of EWSB in this scenario is the 
dimensionful scalar coupling \rfn{si}.

\mysection{DM direct detection and Higgs boson mass}
In our scenario both the DM annihilation at early Universe and the DM scattering on nuclei  
 are dominated by the tree level SM Higgs boson exchange. 
The relevant  DM-Higgs effective coupling  is
\begin{equation}
\lambda_{\text{eff}} \, v = \frac{1}{2} (\sqrt{2} s\, c\, \mu'_{SH} + 2 s^{2} (\lambda_{3}+\lambda_{4}) v +2 c^{2} \lambda_{S1} v),
\label{lameff}
\end{equation}
where $s,\,c$ are the sine, cosine of the singlet-doublet mixing angle. 
Because $\mu'_{SH}\gg \lambda_i v,$ \Eq{lameff} is again dominated by the $\mu'_{SH}$ term.
Therefore the Higgs boson mass from EWSB, the observed DM density of the Universe, and the DM  direct detection
cross section are all related to each other via the soft portal \rfn{si} into DM.

We systematically scan over the full parameter space of the model by iterating between $M_\text{G}$ and $M_h$ using RGEs as described
above. We require successful radiative EWSB and calculate the thermal freeze-out DM abundance and 
spin-independent direct detection cross section per nucleon using MicrOMEGAS package~\cite{micromegas}. 
The latter is approximately given by
\be
 \sigma_{\text{SI}} \approx \frac{1}{\pi} f_N^2 \left( \frac{\lambda_{\text{eff}}v}{v M_{\text{DM}}}  \right)^2 \left( \frac{M_N}{M_h} \right)^4 ,
 \label{sigma}
\ee
where $f_N\approx 0.47$ is the nucleonic form factor and  $M_N$ is the nucleon mass.  
This expression includes all contributions from the valence and sea quarks and gluons.

\begin{figure}[t]
\includegraphics[width=0.4\textwidth]{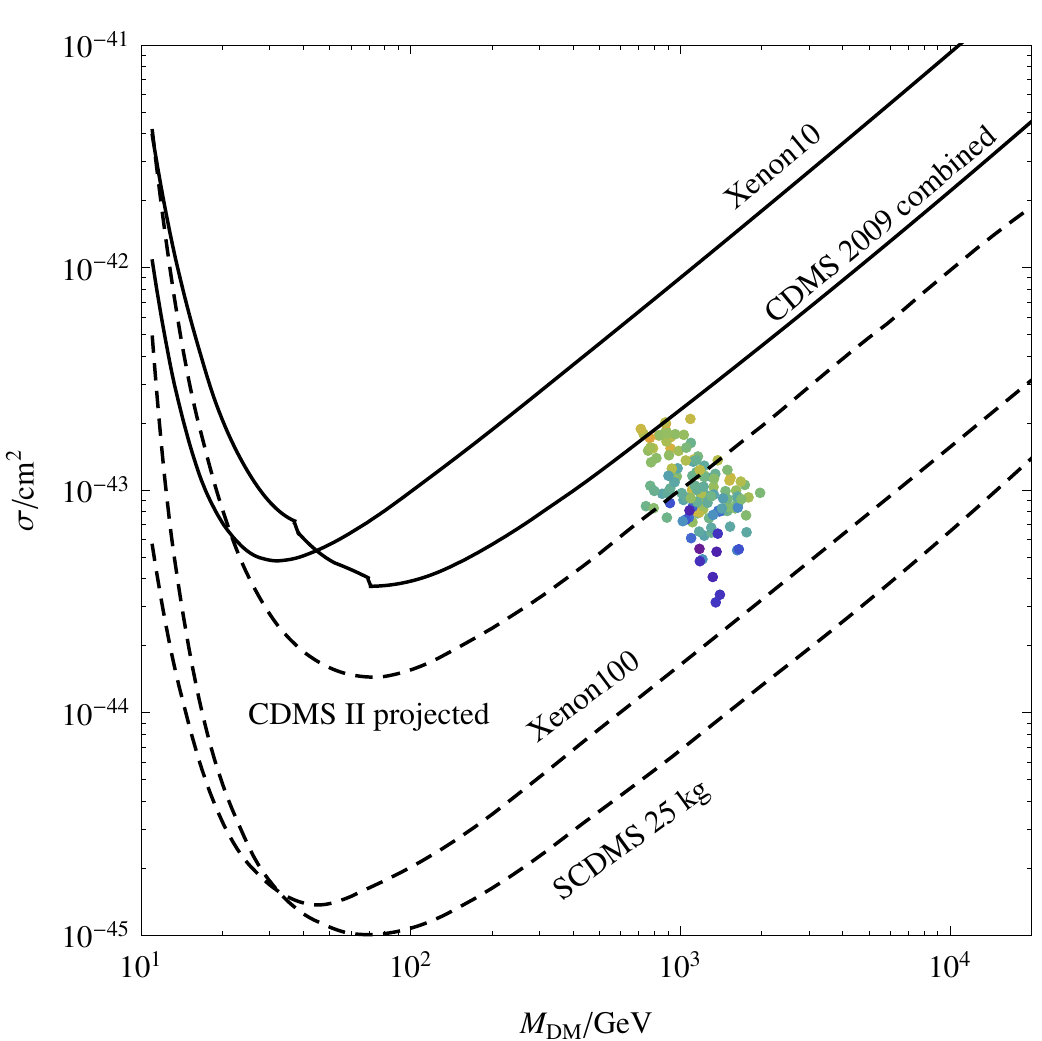}
\caption{ Spin-independent DM direct detection cross-section/nucl.~{\it vs}.~$M_{\mathrm{DM}}$
 for points where $\mu_{1}^{2}$ is negative due to corrections from dark sector. For SM Higgs mass see text.
}
\label{fig3}
\end{figure}

We present in Fig.~\ref{fig3} our prediction for the spin-independent DM cross section as a function of DM mass
for different Higgs boson masses described by the colour code $130~\mrm{GeV}<M_h<180~\mrm{GeV}$ from yellow
to violet. The present experimental bounds (solid lines) \cite{CDMS,Angle:2007uj} together with 
expected sensitivities (dashed lines) are also presented.
For every point we require the WMAP 3$\sigma$ bound $0.094 < \Omega_{\text{DM}} < 0.129.$  Because all the processes are dominated by the soft portal into DM,
we have an amazing prediction for the DM mass range and the spin-independent direct detection cross section $\sigma_{\text{SI}}$ as 
a function of the Higgs mass. The DM mass is restricted into the range $700~\mrm{GeV}<M_{\text{DM}}<2~\mrm{TeV}.$  
The lower bound on $M_{\text{DM}}$ comes  from the requirement of EWSB while the upper bound
is fixed by the DM abundance. 

We predict $\sigma_{\text{SI}}$ just at the present experimental sensitivity. 
CDMS II experiment has observed \cite{CDMS} two candidate events for DM scattering.
Although statistically insignificant at the moment \cite{CDMS}, if confirmed by future experiments, 
the CDMS II result requires light Higgs boson, a prediction that can be tested at LHC.

\mysection{Conclusions}
We have shown that the dimensionful DM coupling \rfn{si} to the SM Higgs boson naturally triggers 
radiative EWSB due to large negative contribution to the SM Higgs mass parameter $\mu^2_1$ in the 1-loop
effective scalar potential. In this scenario EWSB, 
DM thermal freeze-out cross section and DM scattering on nuclei are all dominated by the same coupling.
Therefore we get {\it predictions} both for the DM mass
as well as on its spin independent direct detection cross section, {\it cf.} Fig.~\ref{fig3}. 
Most importantly, for light Higgs boson
the latter is predicted to be just at the present CDMS II sensitivity and can explain the
observed DM scattering events at CDMS II.

%\vskip 0.3 cm

\noindent {\bf Acknowledgment.}
We thank P. Chankowski,  S. Davidson, T. Hambye, K. Huitu, S. Pokorski,  A. Romanino, A. Strumia, M. Tytgat for discussions.
This work was supported by the ESF 8090, JD164 and SF0690030s09.

%\vskip -0.3 cm

\end{document}